\documentclass[12pt]{iopart}

\newcommand{\beq}{\begin{equation}}
\newcommand{\eeq}{\end{equation}}
\newcommand{\bear}{\begin{eqnarray}}
\newcommand{\ear}{\end{eqnarray}}
\newcommand{\nn}{\nonumber \\}

\usepackage{iopams}

\begin{document}

\title{Local spherically symmetric perturbations
of spatially flat Friedmann models}

\author{A.A. Popov}
\address{Department of Mathematics, Kazan State Pedagogical
University, Mezhlauk 1 Street, Kazan 420021, Russia}
\ead{apopov@kzn.ru}
\author{R.K. Muharlyamov}
\address{Department of Physics, Kazan State University,
Kremlevskaja 18 Street, Kazan 420008, Russia}
\ead{Ruslan.Muharlyamov@ksu.ru}

\begin{abstract}
The spherically symmetric perturbations in the spatially flat
Friedman models are considered. It is assumed that
the Friedmannian density and pressure are related through
a linear equation of state. The perturbation
is joined smoothly with an unperturbed Friedmann's
background at the sound horizon of perturbation. Such
junction is in accordance with the "birth" of
a local perturbation as a result of the redistribution
of matter. The solution of the Einstein's equations
is obtained in linear approximation on a Friedmann's
background near the the sound horizon of perturbation.
\end{abstract}


\maketitle

\section{Introduction}

The description of metric perturbations in relativistic cosmology is
significance in the theory of creation of the universe
large-scale structure. Spherically symmetric perturbations are
an important and interesting class of that perturbations.

The study of small perturbations on the cosmological background
has a long history. The first relativistic treatment was given by
Lifshitz \cite{L} and developed by himself and Khalatnikov \cite{LK}.
The gauge-invariant formalism of perturbation was given in
the fundamental paper of Bardeen \cite{3} and was developed
in the works \cite{G}.

Spherically symmetry perturbations on the cosmological background
have been studied by a lot of authors \cite{GS,5,2}.
As it was noted in \cite{2} the spherically symmetric perturbation
in fluid with nonzero pressure may be represent as outgoing and
ingoing waves (travelling from and into the center of
the configuration with the velocity of sound).
The ingoing wave amplitude increases as soon as the
wave approaches to the center of the configuration. Therefore
the magnitude of spherically symmetric perturbation near
the center of configuration must increase  more quickly
than the amplitude of Lifshitz's plane-symmetric wave
perturbations \cite{L}. Moreover the local perturbation
must be smoothly  jointed with an unperturbed Friedmann's
background at the sound horizon of perturbation.

Our main difference from the paper by Ignat'ev and Popov \cite{2}
is that we consider a cosmological fluid with a linear equation
of state  $\varepsilon= \gamma p$, where $\varepsilon$ is density,
$p$ is pressure of fluid but leave arbitrary a constant $\gamma$
($\gamma > 0$).

In Sec. II we write out the rigorous equations describing
the perfect fluid in spherically symmetric space-times and
give the characterization a background space-time.
In Sec. III we fix the gauge and set out the basic equations
governing the spherically symmetric perturbations.
The particular solutions of these equations are given
in Sec. IV. In Sec. V we impose the boundary conditions
on the perturbation and obtain the solution for perturbation
near the boundary. Finally, the main conclusions raised
in this paper are summarized in Sec. VI.

\section{Perfect fluid in spherically symmetric space-time.
Rigorous equations and background space-time}

With the gravitational field being spherically symmetric, coordinates
may be chosen so that the metric takes the form
\begin{equation} \label{1.1}
ds^2=-e^{\nu} d\eta ^2+e^{\lambda} dr^2+ e^{\mu} r^2 \left(
d\theta ^2+{\sin }^2\theta \ d\varphi ^2\right),
\end{equation}
where $\nu $, $\mu$ and $\lambda$ are functions of the time
coordinate $\eta$ and a radial space coordinate $r$.
As we are interested in perfect-fluid distribution,
the energy-momentum tensor is
\beq
T^{\mu}_{\nu}=(\varepsilon +p ) u^{\mu} u_{\nu} +p
\delta^{\mu}_{\nu},
\eeq
where $\varepsilon$ is density, $p$ is pressure and
$u^{\mu}(u^{\eta},u^r,0,0)$ is four-velocity of fluid.
Take into account the equation of state
\beq  \label{ep}
\varepsilon=\gamma p
\eeq
one can obtain the following nontrivial Einstein equations
$(c=G=1)$
\bear \label{1.4}
e^{-\lambda }\left( \mu'' +\frac {3 \mu'^2}{4}
+\frac{3 \mu'}{r} -\frac{\mu' \lambda'}{2}- \frac{\lambda'}{r} +
\frac{1}{r^2} \right) &-&\frac{e^{-\mu }}{r^2} -e^{-\nu
}\left(\frac{{\dot \mu}^2}{4} +\frac{\dot \mu \dot
\lambda}{2}\right)\nn &=& -8\pi p\left[ \gamma+(1+\gamma)\upsilon
^2\right],
\ear
\bear \label{1.2}
e^{-\lambda } \left( \frac{
\mu'^2}{4} +\frac{\mu' \nu'}{2} +\frac{\mu' +\nu'}{r}
+\frac{1}{r^2}\right) -\frac{e^{-\mu}}{r^2}
&+&e^{-\nu}\left(-\ddot \mu -\frac {3{\dot \mu}^2}{4} +\frac{\dot
\mu \dot \nu}{2} \right)\nn &=& 8 \pi p\left[ 1+(1+\gamma)\upsilon
^2\right],
\ear
\bear \label{1.3}
e^{-\lambda } &&  \left[
\frac{\mu''+\nu''}{2} +\frac{ \mu'^2+\nu'^2 +\mu' \nu'-\mu'
\lambda'-\nu' \lambda' }{4} +\frac{2 \mu' +\nu'
-\lambda'}{2 r} \right]  \nn
&&+e^{-\nu} \left[ -\frac{(\ddot \mu +\ddot
\nu)}{2} +\frac{{\dot \nu} \dot \lambda
+\dot \nu \dot \mu  -{\dot\mu}^2 -\dot \mu \dot \lambda -{\dot
\lambda}^2 }{4}\right] =8 \pi p,
\ear
\begin{equation} \label{1.5}
e^{-\nu }\left[{\dot \mu}'+\frac{\mu' \dot \mu}{2} -\frac{(\mu'
\dot \lambda +\dot \mu \nu')}{2} +\frac{\dot \mu -\dot \lambda}{r}
\right] =8\pi e^{\left( \lambda -\nu \right) /2}
(1+\gamma)p\upsilon \sqrt{1+\upsilon ^2},
\end{equation}
where $\upsilon=u^re^{\lambda /2}$ is the frame projection of radial
velocity, a dot and a prime denote partial derivatives with respect to
$\eta$ and $r$, respectively.

In isotropic spherical coordinates the metric of the background
space-time can be written as
\begin{equation} \label{2.1}
ds^2=a^2\left\{ -d\eta ^2+ dr^2+r^2\left( d\theta
^2+\sin {}^2\theta \ d\varphi ^2\right) \right\},
\end{equation}
where the function $a(\eta)$ satisfy the equation
\beq \label{fon}
2 \gamma \frac{\ddot a}{a}=\left( \gamma -3
\right)\frac{{\dot a}^2}{a^2}.
\eeq
The energy density $\varepsilon_0$, the pressure $p_0$ and the frame
projection of radial velocity of the background fluid $\upsilon _0$ are
\begin{equation}
\label{2.4}\varepsilon_0=\frac{3\stackrel{.}{a}^2}{8\pi a^4},
\quad \varepsilon_0=\gamma p_0,\quad v_0=0.
\end{equation}

\section{Gauge transformation and perturbed equations}

We shall only consider spherically symmetric perturbations of
spatially flat Friedmann models, i.e.
\begin{equation} \label{nu}
\nu=\ln \left( a^2\right)+\delta \nu, \ \lambda= \ln \left(
a^2\right)+\delta \lambda, \ \mu= \ln \left( a^2\right)+\delta
\mu,
\end{equation}
\begin{equation} \label{lambda}
\varepsilon=\varepsilon_0+\delta \varepsilon, \ p=p_0+\delta p,
\end{equation}
where $\delta \nu $, $\delta \lambda$, $\delta \mu$, $\delta
\varepsilon$ $\delta p$ are the functions of the coordinates
$\eta $ è $r$ such that
\beq
|\delta \nu|, |\delta \lambda|, |\delta \mu| \ll 1,
 \ \delta \varepsilon \ll \varepsilon_0, \ \delta
p \ll p_0.
\eeq
We shall assume that the frame projection of radial velocity of
fluid $\upsilon $ is the small quantity too.

Let us note that the linearized gauge transformation
\beq
\eta=\tilde \eta +\delta \eta (\tilde \eta, \ \tilde r), \
r=\tilde r +\delta r(\tilde \eta, \ \tilde r)
\eeq
gives
\bear
ds^2&=&-a^2(\eta) (1+\delta \nu) d \eta^2 +a^2(\eta)
(1+\delta \lambda) d r^2 +a^2(\eta) r^2 (1+\delta \mu)
d \Omega^2 \nn
&=&-a^2(\tilde \eta) \left(1+\delta \nu +2\delta \dot \eta
+2\frac{\dot a}{a} \delta \eta \right) d \tilde {\eta}^2
+2 a^2(\tilde \eta) (\delta \dot r +\delta \eta ') d\tilde{\eta} d\tilde r^2
\nn &&+a^2(\tilde \eta) \left(1 +\delta \lambda +2\delta r'
+2\frac{\dot a}{a} \delta \eta \right) d \tilde{r}^2
\nn &&
+a^2(\tilde \eta) \tilde{r}^2 \left(1+\delta \mu
+2 \frac{\delta r}{\tilde{r}}+2\frac{\dot a}{a} \delta \eta \right)
d \Omega^2
\ear
Here the dot and the prime denote $\partial/\partial \tilde \eta$ and
$\partial/\partial \tilde r$ respectively,
$d \Omega^2 = d\theta ^2 +{\sin }^2\theta \ d\varphi ^2$.
The functions $\delta \eta$ and $\delta r$ can be chosen such that
\beq
\frac{d (\delta r)}{d \tilde \eta} + \frac{d (\delta \eta)}{d \tilde r}=0, \
\frac{d (\delta r)}{d \tilde r} -\frac{\delta
r}{\tilde r}= \frac{\delta \mu-\delta \lambda}{2}
\eeq
and consequently $\delta \tilde{ \lambda}=\delta \tilde{\mu}$.
Omitting the tilde over $\eta, \ r, \ \delta \lambda$ and $\delta {\mu}$ below
we  shall suppose
\beq \label{st}
\delta {\mu}=\delta \lambda.
\eeq
Now one can to rewrite the metric in linear (with respect
to the perturbations) approximation as
\beq \label{is}
ds^2=-a^2(\eta) \left\{(1+\delta \nu) d \eta^2 +
(1+\delta \lambda)\left[d r^2 + r^2 (1+\delta \mu)
d \Omega^2\right]\right\}
\eeq

Thus Einstein's equations (\ref{1.4}-\ref{1.5}) are
reduced in linear approximation to the following four equations
\beq \label{01} {\delta \lambda}'' +2\frac{{\delta \lambda}'}{r}
-3\frac{\dot a}{a} \delta \dot {\lambda} +3\frac{{\dot a}^2 }{a^2}
\delta \nu =-8 \pi a^2 \delta \varepsilon, \eeq \beq \label{02}
\frac{{\delta \nu}'+{\delta \lambda}'}{r} -\delta \ddot \lambda
-2\frac{ \dot a }{a}\delta \dot {\lambda} +\frac{{\dot a}}{a}
\delta \dot {\nu} +\left( 2\frac{\ddot a}{a}-\frac{{\dot
a}^2}{a^2} \right)\delta \nu =8 \pi a^2 \frac{\delta
\varepsilon}{\gamma}, \eeq \beq \label{03} \frac{{\delta
\nu}''+{\delta \lambda}''}{2} +\frac{{\delta \nu}'+ {\delta
\lambda}'}{2 r} - \delta {\ddot \lambda} -2\frac{ \dot a }{a}
{\delta \dot \lambda} +\frac{{\dot a}}{a} {\delta \dot \nu}
+\left( 2\frac{\ddot a}{a}-\frac{{\dot a}^2}{a^2} \right)\delta
\nu =8 \pi a^2 \frac{\delta \varepsilon}{\gamma},
\eeq
\beq \label{04}
{\delta \dot \lambda}' -\frac{\dot a}{a} {\delta \nu}'
=8 \pi \left(1+\frac{1}{\gamma}\right) a^2 \varepsilon_0 \upsilon
\eeq
for the four unknown functions $\delta \nu, \delta
\lambda, \delta \varepsilon, \upsilon$ (we take into account
$\delta p=\delta \varepsilon / \gamma $).

Let us note that as the consequence of the symmetry of problem
the perturbations described in this paper are scalar ones.
The gauge-invariant amplitude of density perturbation with the notations of
the paper \cite{3} is
\begin{equation}
\label{3.22}\epsilon_g=\delta - 3(1+ \mathop{\rm w}) \frac 1{\mathop{\rm k}}
\frac{\dot S}S \left( B^{(0)} - \frac 1{\mathop{\rm k}} \dot H^{(0)}_T
\right)
\end{equation}
The choice of the isotropic coordinates (\ref{is}) in our paper corresponds
to a longitudinal gauge. With the notations of the paper \cite {3} this give
\begin{equation}
\label{3.23}B^{(0)} = H^{(0)}_T = 0.
\end{equation}
By comparing the corresponding expressions in two papers we find
\begin{equation}
\label{3.24} \epsilon_g Q^{(0)}=\delta Q^{(0)}=\delta \varepsilon /
\varepsilon_0 .
\end{equation}
The relations between the other perturbed quantities can be determined
analogously
\begin{equation}
\label{3.25}v^{(0)}_s Q^{(0)}=\left[ v^{(0)} - \frac1{\mathop{\rm k}}
\dot H^{(0)}_T \right] Q^{(0)}=v^{(0)} Q^{(0)}=v,
\end{equation}
\bear \label{3.26}
\Phi_A Q^{(0)}&=&\left[ A+\frac 1{\mathop{\rm k}}
\dot B^{(0)} + \frac 1{\mathop{\rm k}} \frac {\dot S}S B^{(0)} -
\frac 1{{\mathop{\rm k}}^2} \left( \ddot H^{(0)}_T +
\frac{\dot S}S \dot H^{(0)}_T \right) \right] Q^{(0)}
\nn &=&A Q^{(0)}=\frac{\delta \nu}2,
\ear
\begin{equation}
\label{3.27} \Phi_H Q^{(0)}=\left[ H_L + \frac13 H^{(0)} + \frac1
{\mathop{\rm k}} \frac{\dot S}S B^{(0)} - \frac 1{\mathop{\rm k}^2}
\frac{\dot S}S \dot H^{(0)}_T \right] Q^{(0)}=H_L Q^{(0)}=
\frac{\delta \lambda}2.
\end{equation}
From this relations we find that the perturbation quantities introduced
in this section are actualy physical as well as the gauge-invariant
quantities $\epsilon_g Q^{(0)}, v^{(0)}_s Q^{(0)}, \Phi_A Q^{(0)},
\Phi_H Q^{(0)}$.

The difference of equations (\ref{02}) è (\ref{03}) gives
\begin{equation} \label{3.8}
\left( \delta \lambda''+\delta \nu''\right)= \frac 1r\left( \delta
\lambda' +\delta \nu'\right).
\end{equation}
The solution of this equation is
\begin{equation}
\label{3.8a}\delta \lambda +\delta \nu =C_1+C_2r,
\end{equation}
where $C_1$ and $C_2$ are arbitrary functions of the time
coordinate $\eta$. If we require that
\beq
\delta \lambda= \delta \nu =0, \qquad \mbox{for} \ r \rightarrow \infty,
\eeq
i.e. the space-time is asymptotically homogeneous and isotropic, then
\beq
C_1=C_2=0
\eeq
and
\begin{equation}
\label{3.8b}
\delta \nu= - \delta \lambda.
\end{equation}
One can to consider the expressions (\ref{01}) and (\ref{04})
as the definitions of $\delta \varepsilon$ and $\upsilon$
respectively
\begin{equation} \label{pe}
8\pi \delta \varepsilon=8\pi \gamma\delta p=
\frac{3\stackrel{.}{a}}{a^3}\left( \delta \stackrel{.}{\lambda
}+\frac{\stackrel{.}{a}}a\delta \lambda \right) -
\frac{1}{a^2}\left[ \delta \lambda ^{\prime \prime }+ \frac 2r
\delta \lambda ^{\prime } \right],
\end{equation}
\begin{equation} \label{3.18}
8\pi \upsilon =\frac{\gamma}{a^2 (1+\gamma)
\varepsilon _0}\left( \delta \stackrel{.}{\lambda }^{\prime
}+\frac{\dot a}{a} \delta \lambda ^{\prime}\right).
\end{equation}
Thus the spherically symmetric perturbations can be described
by the metric
\beq
ds^2=-a^2(\eta) \left\{(1-\delta \lambda) d \eta^2 +
(1+\delta \lambda)\left[d r^2 + r^2 (1+\delta \mu)
d \Omega^2\right]\right\},
\eeq
where the single unknown function $\delta \lambda (\eta, r)$
is determined by the following combination
of the expressions (\ref{01}-\ref{03})
\begin{equation} \label{3.9}
\delta \lambda'' +\frac{2}{r}\delta \lambda'= \gamma\delta \ddot
\lambda+3 \left( 1+\gamma \right) \frac{\dot a}{a}\delta \dot
\lambda.
\end{equation}

\section{Particular solutions}

First of all it is necessary to note that the equation
(\ref{3.9}) has a solution, which correspond to the Newtonian
potential \cite{mcv}
\beq
\delta \lambda =- \delta \nu = \frac{2 m}{a r},
\eeq
caused by a particle of variable mass
\beq
m(\eta)=C_3 a +C_4 a^{-3(1+\gamma)/(2 \gamma)},
\eeq
where $C_3$ and $C_4$ are constants.

If the  perfect fluid is a dust, i.e.
\beq
p=0 \quad (1/\gamma = 0),
\eeq
then the metric perturbations are described by the equation
\begin{equation} \label{10.2}
\frac{\partial^2 (\delta \lambda)}{\partial a^2}
+\frac{7 }{2 a}\frac{\partial (\delta \lambda)}{\partial a} =0.
\end{equation}
The solution of this equation is
\beq
\delta \lambda = \frac{F_1}{a^{5/2}}+F_2,
\eeq
where $F_1$ and $F_2$  are the arbitrary functions of $r$.

The case of the ultra relativistic fluid ($\varepsilon=3 p$)
has been considered in \cite{2}. The corresponding solution
has a form
\beq
\delta \lambda= \frac{1}{a r} \frac{ \partial}{\partial \eta}
\left\{ \frac{1}{a} \left[ \Phi_+\left(\frac{\eta}{\sqrt{3}}+r\right)
+\Phi_-\left(\frac{\eta}{\sqrt{3}}-r\right) \right] \right\},
\eeq
where $\Phi_+(x)$ è $\Phi_-(x)$ are the arbitrary functions.

\section{Boundary conditions and solution near the sound horizon}

The hyperbolic equation (\ref{3.9}) describes the outgoing and
ingoing waves travelling with speed of sound from and into
the center of the configuration. Therefore if we like to describe
a local perturbation of radius $r_0$ (at the moment $\eta_0$)
we can impose the boundary
conditions at the sound horizon of this perturbation
\begin{equation} \label{5.6}
\left. \delta \lambda \right| _{r-r_0-\tau +\tau _0=0}
=\left. \frac{ \partial (\delta \lambda) } {\partial r}\right| _
{r-r_0-\tau +\tau _0=0}=\left. \frac{ \partial (\delta \lambda) }
 {\partial \eta} \right| _{r-r_0-\tau +\tau _0=0}=0,
\end{equation}
where
\beq
\tau=\eta/\sqrt{\gamma}.
\eeq
Such boundary conditions are in accordance with the "birth" of
a perturbation as a result of the redistribution of matter.
If we make the substitution
\beq \label{dl}
\delta \lambda = \frac{V(\tau, r)}{r a^{\beta}},
\eeq
where
\beq \label{bt}
\beta=\frac32 \left( \frac{1+\gamma}{\gamma}\right),
\eeq
and introduce new coordinates
\bear
x&=&\tau+r, \nn y&=&\tau-r-\tau_0+r_0,
\ear
then Eq. (\ref{3.9}) can be rewritten in the form
\beq \label{eqV}
4 \frac{\partial^2 V(x,y)}{\partial x \partial y}
-\beta \left(  \frac{d a}{a d \tau}_{|\tau=(x+y+\tau_0-r_0)/2}
\right)^2 V(x,y)=0.
\eeq
We shall find the solution of this equation near the boundary
$y=0$ representing the functions
$ \left({d a}/(a d \tau)\right)_{|\tau=(x+y+\tau_0-r_0)/2}$
and  $V(x,y)$ as a power series
\beq
\frac{d a}{a d \tau}_{|\tau=(x+y+\tau_0-r_0)/2}=
\sum_{k=0}^{\infty}\frac{y^k}{k!}\left[\left( \frac{\partial}
{2 \partial \tau} \right)^k \frac{d a}{a d \tau}\right]_{|\tau=(x+\tau_0-r_0)/2},
\eeq
\beq \label{serV}
V(x,y)=\Theta(y)\sum_{n=2}^{\infty} V_n(x) \ y^n.
\eeq
Here we take into account that the first terms of last series
vanish as a consequence of the boundary conditions (\ref{5.6}).
$\Theta$-function
\beq
\Theta(y)=\left\{\begin{array}{l}{0, \quad y < 0,}\\
{1, \quad y \geq 0} \end{array} \right.
\eeq
is introduced into expression (\ref{serV}) since the perturbation
must be vanish out of the sound horizon.
Then the equation (\ref{eqV}) is rewritten as
\bear
&&\sum_{n=2}^{\infty}\left\{ 4 n \frac{d V_n}{d x} y^{n-1}
-\beta V_n \sum_{k,s=0}^{\infty}\frac{y^{k+s+n}}{k! s!}
\left[\left( \frac{\partial}{2 \partial \tau} \right)^k
\frac{d a}{a d \tau}\right]
\right. \nn && \left.
\left[\left( \frac{\partial}{2 \partial \tau} \right)^s
\frac{d a}{a d \tau}\right]_{|\tau=(x+\tau_0-r_0)/2}
\right\}=0.
\ear
The first two terms of this series give
\bear
\frac{d V_2}{d x}&=&0, \nn
\frac{d V_3}{d x}&-&\frac{\beta}{12} V_2 \left( \frac{d a}{a d \tau}
 \right)^2 _{|\tau=(x+\tau_0-r_0)/2}=0.
\ear
The solutions of these equations are
\bear
V_2&=&\Psi_2, \nn
V_3&=&\Psi_3+\frac{\Psi_2 \beta}{6(1-\beta)}
\left( \frac{d a}{a d \tau}  \right) _{|\tau=(x+\tau_0-r_0)/2}.
\ear
Here we take into account that
\beq
\int dx \left( \frac{d a}{a d \tau}  \right)^2 _{|\tau=(x+\tau_0-r_0)/2}=
2\left[\int d\tau \left( \frac{d a}{a d \tau}  \right)^2
\right] _{|\tau=(x+\tau_0-r_0)/2}
\eeq
and
\bear
\int d\tau \left( \frac{d a}{a d \tau}  \right)^2 &=&
\beta \int d\tau \left( \frac{d a}{a d \tau}  \right)^2+
\frac{d a}{ad \tau}
\ear
Thus
\bear
V(x,y)&=&\left\{\Psi_2+\left[\Psi_3+ \frac{\Psi_2 \beta}{6(1-\beta)}
\left( \frac{d a}{a d \tau}  \right) _{|\tau=(x+\tau_0-r_0)/2}
\right] y \right. \nn && \left.
+O\left( \frac{y^2}{(x+\tau_0-r_0)^2} \right)
\right\}\Theta(y).
\ear
The coefficient $1/(r a^{\beta})$ in (\ref{dl}) must be expand
in series of $y$ also
\bear
\frac{1}{r a^{\beta}(\tau)}&=&\frac{2}{(x-\tau_0+r_0)
a^{\beta}_{|\tau=(x+\tau_0-r_0)/2}}
\left\{ 1+ \left[\frac{1}{(x-\tau_0+r_0)}\right. \right. \nn &&
\left. \left. -\frac{\beta}{2}
\left( \frac{d a}{a d \tau}  \right) _{|\tau=(x+\tau_0-r_0)/2}
\right]y + O\left( \frac{y^2}{(x+\tau_0-r_0)^2} \right) \right\}.
\ear
Thus
\bear
\delta \lambda(x,y) &=&\frac{2 y^2}{(x-\tau_0+r_0)
a^{\beta}_{|\tau=(x+\tau_0-r_0)/2}}
\left\{ \Psi_2+ \left[ \Psi_3+\frac{\Psi_2}{(x-\tau_0+r_0)}
\right. \right. \nn && \left. \left.
+\Psi_2 \frac{\beta(3 \beta-2)}{6(1-\beta)}
\left( \frac{d a}{a d \tau}  \right) _{|\tau=(x+\tau_0-r_0)/2}
\right]y
\right. \nn && \left.
+ O\left( \frac{y^2}{(x+\tau_0-r_0)^2} \right) \right\} \Theta(y),
\ear
or, if we take into account (\ref{bt})
\bear \label{fin}
\delta \lambda(\tau,r) &=&\frac{ (\tau-r-\tau_0+r_0)^2}{
(\tau+r-\tau_0+r_0) a^{\beta}(z)}
\left\{\Psi_2+
\left[
\Psi_3+\frac{\Psi_2}{(\tau+r-\tau_0+r_0)}
\right. \right.   \nn &&   \left. \left.
-\Psi_2 \frac{(9+14 \gamma +5\gamma^2)}{4 \gamma(\gamma+3) a(z)}
\frac{d a(z)}{d z}
\right](\tau-r-\tau_0+r_0)
\right. \nn && \left.
+ O \left( \frac{(\tau-r-\tau_0+r_0)^2}{(\tau+r+\tau_0-r_0)^2}
\right) \right\}_{| z=(\tau+r+\tau_0-r_0)/2}
\nn &&
{\Theta \left( \tau-r-\tau_0+r_0 \right)}.
\ear
For the case $\Psi_3=\tau_0=r_0=0$, $\gamma=3$, this expression
coincides with the expansion of $\delta \lambda$ from \cite{5}  near
the sound horizon.

\section{Conclusion}

We have considered the spherically symmetric perturbations
of a perfect fluid on the background of spatially flat Friedmann
models. It was assumed that the fluid is described by the linear
equation of state (\ref{ep}). It was shown that in longitudinal
gauge the local perturbations
of the energy density, the pressure, the frame projection of
the radial velocity of fluid (\ref{pe}, \ref{3.18}) and the metric
coefficients (\ref{st}, \ref{3.8b}) are determined by the single
function. This function satisfies equation (\ref{3.9}) of hyperbolic
type and describes the outgoing and ingoing waves travelling with speed
of sound from and into the center of the configuration.
The boundary conditions (\ref{5.6}) on the solution of this equation
are imposed at the sound horizon of perturbation. The solution
is represented as a power series (\ref{dl}, \ref{serV}) near
this horizon. First two terms of series are obtained obviously
(\ref{fin}). The corresponding to these terms solution coincides with
the expansion near sound horizon early known one in the case
of ultrarelativistic equation of state ($\varepsilon=3 p$).

\ack
We would like to thank S.V. Sushkov  N.R. Khusnutdinov for helpful
conversations.
This work was supported in part by grants 05-02-17344, 05-02-39023
and 06-01-00765 from the Russian Foundation for Basic Research.

\end{document}